\documentclass[fleqn,10pt]{wlscirep}
\title{Near-infrared laser thermal conjunctivoplasty}

\author[1]{Jianlong Yang}
\author[1]{Rahul Chandwani}
\author[1]{Varun Gopinatth}
\author[1]{Tim Boyce}
\author[2]{Stephen C. Pflugfelder}
\author[1]{David Huang}
\author[1,*]{Gangjun Liu}
\affil[1]{Casey Eye Institute, Oregon Health and Science University,3375 SW Terwilliger Blvd, Portland, Oregon,97239, USA}
\affil[2]{Ocular Surface Center, Cullen Eye Institute, Baylor College of Medicine, 6565 Fannin NC205, Houston, TX 77030, USA}

\affil[*]{liga@ohsu.edu}


\begin{abstract}
Conjunctivochalasis is a common cause of tear dysfunction due to the conjunctiva becoming loose and wrinkly with age. The current solutions to this disease include either surgical excision in the operating room, or thermoreduction of the loose tissue with hot wire in the clinic. We developed a near-infrared (NIR) laser thermal conjunctivoplasty (LTC) system. The system utilizes a 1460-nm programmable laser diode system as the light source. At this wavelength, a water absorption peak exists and the blood absorption is minimal, so the heating of redundant conjunctiva is even and there is no bleeding. A miniaturized handheld probe delivers the laser light and reshapes the laser into a 10$\times$1 mm$^2$ line on the working plane. A foot pedal is used to deliver a preset number of calibrated laser pulses. A fold of loose conjunctiva is grasped by a pair of forceps.  The NIR laser light is delivered through an optical fiber and a laser line is aimed exactly on the conjunctival fold by a cylindrical lens. Ex vivo experiments using porcine eye was performed to investigate the shrinkage of conjunctiva induced by the NIR laser and decide the optimal laser parameters. It was found that up to 50\% of conjunctiva shrinkage could be achieved. 
\end{abstract}
\begin{document}

\flushbottom
\maketitle
% * <john.hammersley@gmail.com> 2015-02-09T12:07:31.197Z:
%
%  Click the title above to edit the author information and abstract
%
\thispagestyle{empty}

%\noindent Please note: Abbreviations should be introduced at the first mention in the main text – no abbreviations lists. Suggested structure of main text (not enforced) is provided below.

\section*{Introduction}

The sclera, white part of the eye, is covered by a thin, clear membrane called the conjunctiva.  Like skin, the conjunctiva becomes loose and wrinkly with age; this degenerative condition is called conjunctivochalasis \cite{meller98}. The loose folds of conjunctiva often disrupt the uniform distribution of tears and can cause constant eye irritation and blurred vision \cite{liu86,di04}. In severe cases, the conjunctival folds protrude onto the inferior eye lid margin and are traumatized by the lid during blinking.  Furthermore, the lid’s skin is also irritated and altered by the displaced tear.\\
\indent Conjunctivochalasis is a common cause of tear dysfunction (``dry eyes''); however, it does not respond to the usual dry eye treatments such as artificial tears, punctal plugs and anti-inflammatory drops. Effective treatment requires surgical reduction or excision of the redundant conjunctiva to reestablish the inferior tear meniscus and normal tear dynamics. Conjunctivochalasis is typically diagnosed by evaluating the conjunctiva for redundant folds that prolapse onto the lower eyelid and obliterate the tear meniscus in that region. Cross-sectional optical coherence tomography (OCT) allows us for evaluating the severity of the condition and the effectiveness of surgery to remove the redundant tissue \cite{zhang13,zhang11,gumus10}.\\
\indent Surgical means used to remove redundant conjunctival tissue is an effective way to treat conjunctivochalasis. However, the current surgical techniques, such as thermocautery or electrocautery, are not performed on a widespread basis due to the long painful healing period \cite{nak12,ohba11,oh99}.  Both techniques reach temperatures exceeding the boiling point of water and can burn the conjunctival epithelium, underlying stroma, and surrounding tissue. This creates a full thickness burn wound that is generally painful, takes up to one month to fully heal, and occasionally induces excessive inflammation with scarring. Furthermore, a chronic inflammatory conjunctival mass called pyogenic granuloma could result, which would necessitate long-term anti-inflammatory eye drops and possibly further surgery. Poor cosmetic appearance (i.e. red blots due to bleeding in surface tissue) or scarring during the long healing period also deters patients from these techniques. Surgical conjunctival excision with the addition of an amniotic membrane transplant (attached by fibrin glue or suture) can improve the healing course, but must be performed in the operating room, which markedly increases cost.  Laser conjunctivoplasty, on the other hand, can well constrain the thermal energy in the target area and depth and precisely control the amount of energy thus avoiding burning the conjunctival epithelium and underlying stroma.\\
\indent Argon lasers have been used for laser conjunctivoplasty \cite{kumar03,naval95}. However, they work at a wavelength of 532 nm and the laser energy is primarily absorbed by blood and can lead to vessel rupture and hemorrhage. Because the water content of conjunctiva is around 70\% \cite{ferrari09}, heating the conjunctiva using water absorption is a more effective and safe way to achieve shrinkage. In this paper, we proposed to use the near-infrared (NIR) light for LTC, because the blood absorption at this waveband is much weaker than that at visible region \cite{kuen95,horecker43} while two strong water absorption peaks exist at 1460 nm and 1940 nm \cite{curcio51}. Compared with the expensive and inefficient lasers at 1940 nm, the diode laser at 1460 nm has an electro-optical efficiency of $>$ 40\% and the price of the commercial products is reasonable. Thus we developed a miniature-probed-based LTC system using a laser diode system at this wavelength. Ex vivo experiments on porcine eye were conducted to demonstrate the effectiveness of the NIR LTC.
\section*{Results}

\subsection*{Temperature changes in time domain}
\begin{figure}[!htb]
\centering
\includegraphics[width=13cm]{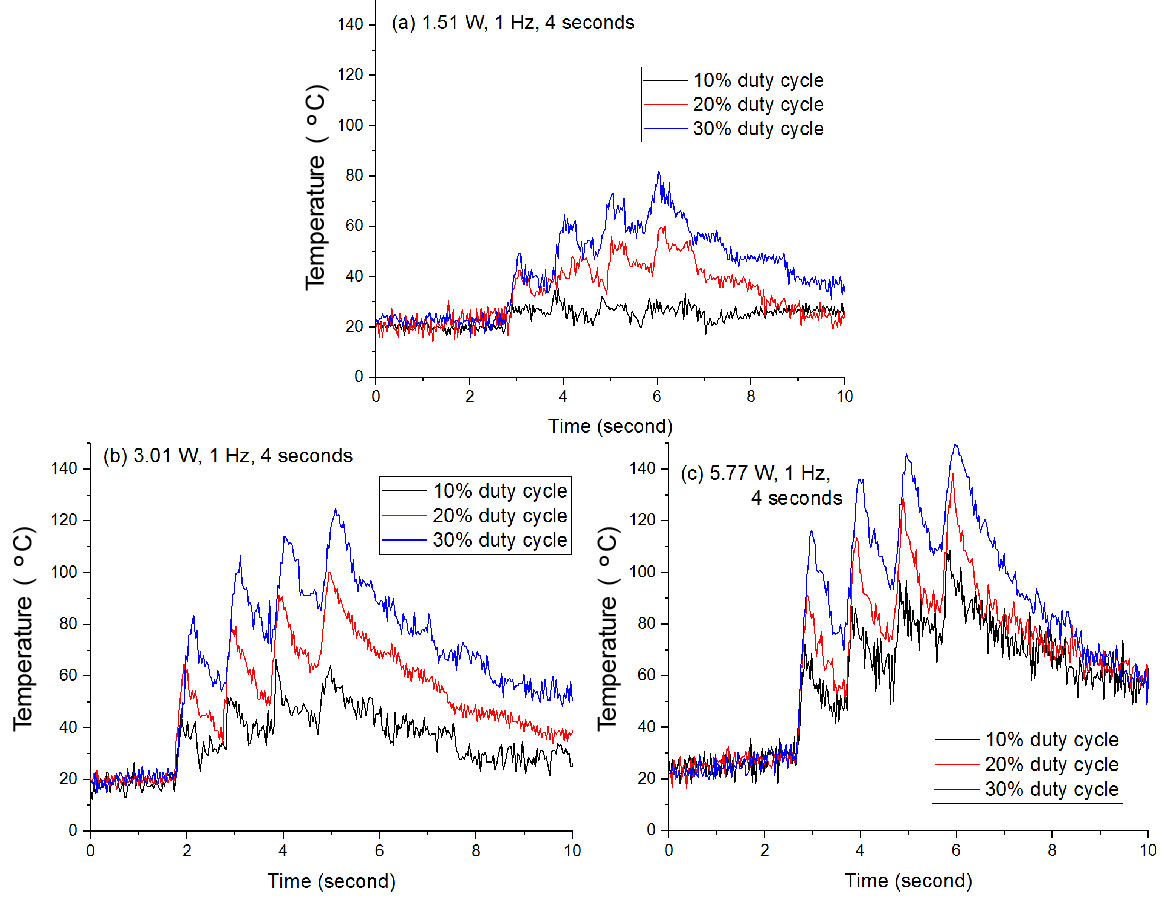}
\caption{Temperature changes during the LTC process using different laser parameters.}
\end{figure}
During the laser heating procedure, the temperature continues to increase until the laser is turned off. However, to optimize the shrinking, the temperature should be kept in a range to avoid burning and boiling. The collagen shrinkage temperate is around 62 $^{\circ}$C \cite{rigby60} and to avoild boiling, the temperature should be kept below (100 $^{\circ}$C). So idea temperature is between (62 $^{\circ}$C)  and (100 $^{\circ}$C).   In order to achieve this, we use pulse laser for the heating and  a pulse repetition rate of 1 Hz and a working duration of 4 seconds (a total of 4 pulses) were selected.     Figure 1 shows the temperature changes during the LTC. The highest temperatures in each of recorded thermal camera image sequences were obtained. It can be seen that the peak temperature increases very fast during the laser pulse illumination period and decreases relatively slowly during the laser-off period. The effects of peak laser power and pulse duty cycle were investigated. With the laser peak power of 1.51 W and duty cycle of 10\% (100 ms pulse duration), the temperature increase is very small and the temperature is always below 40 $^{\circ}$C [Fig. 1(a)]. When the duty cycle was increased to 20\% (200 ms pulse duration) and 30\% (300 ms pulse duration), the cumulative effect of multiple pulses can be clearly seen.  The temperature will increase during the laser-illumination period and the temperature will drop during the laser-off period.  This temperature increase-decrease process will continue with the increase of the pulse numbers. For a peak power of 3 W, the maximum temperature for the 30\% duty cycle exceeds 100 $^{\circ}$C while the maximum temperature for the 10\% duty cycle is around 65 $^{\circ}$C [Fig. 1(b)]. When the peak power was further increased to 5.77 W, the maximum temperature exceeds 100 $^{\circ}$C for all three different duty cycles [Fig. 1(c)].  

\subsection*{OCT imaging}
 \begin{figure}[!htb]
\centering
\includegraphics[width=11cm]{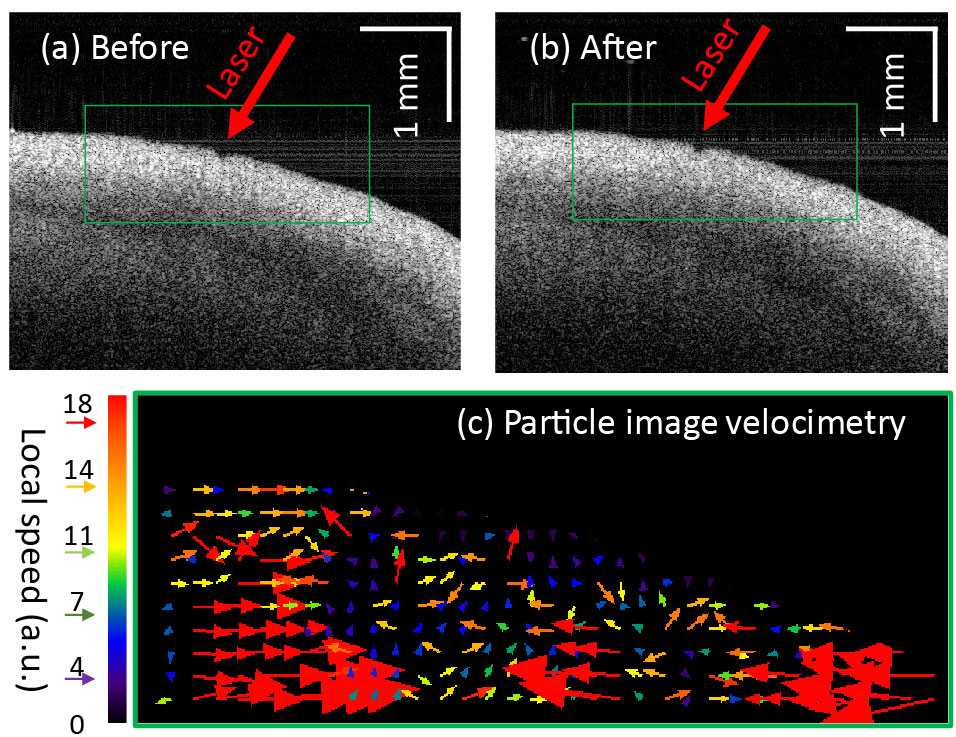}
\caption{OCT B-frames taken before (a) and after (b) the LTC process (A video clip of the time-sequence OCT B-scans can be found in Supplementary information).  (c) PIV images resulting from the green boxes in (a) and (b). The different colors correspond to different speeds and the arrows indicate the moving direction.}
\end{figure}
Real-time OCT imaging was used to monitor the LTC-induced shrinkage process. This dynamic process can be clearly visualized in a video of the time-sequence OCT B-scans (Media 1), but is difficult to see from the OCT images before [Fig.2 (a)] and after [Fig. 2(b)] the LTC. To further visualize and quantify the shrinkage from these images, particle image velocimetry (PIV) was employed \cite{adrian91}. The green boxes in Fig. 2(a) and (b) were employed to calculate the PIV using ImageJ \cite{sch12}. The results are shown in Fig. 2(c). The different colors correspond to different speeds and the arrows indicate the moving direction. The red arrow groups at both sides point to the central region, where was heated by the laser. This is a clear indication of the shrinkage process.
\subsection*{LTC induced shrinkage and optimized laser parameters}
 \begin{figure}[!htb]
\centering
\includegraphics[width=13cm]{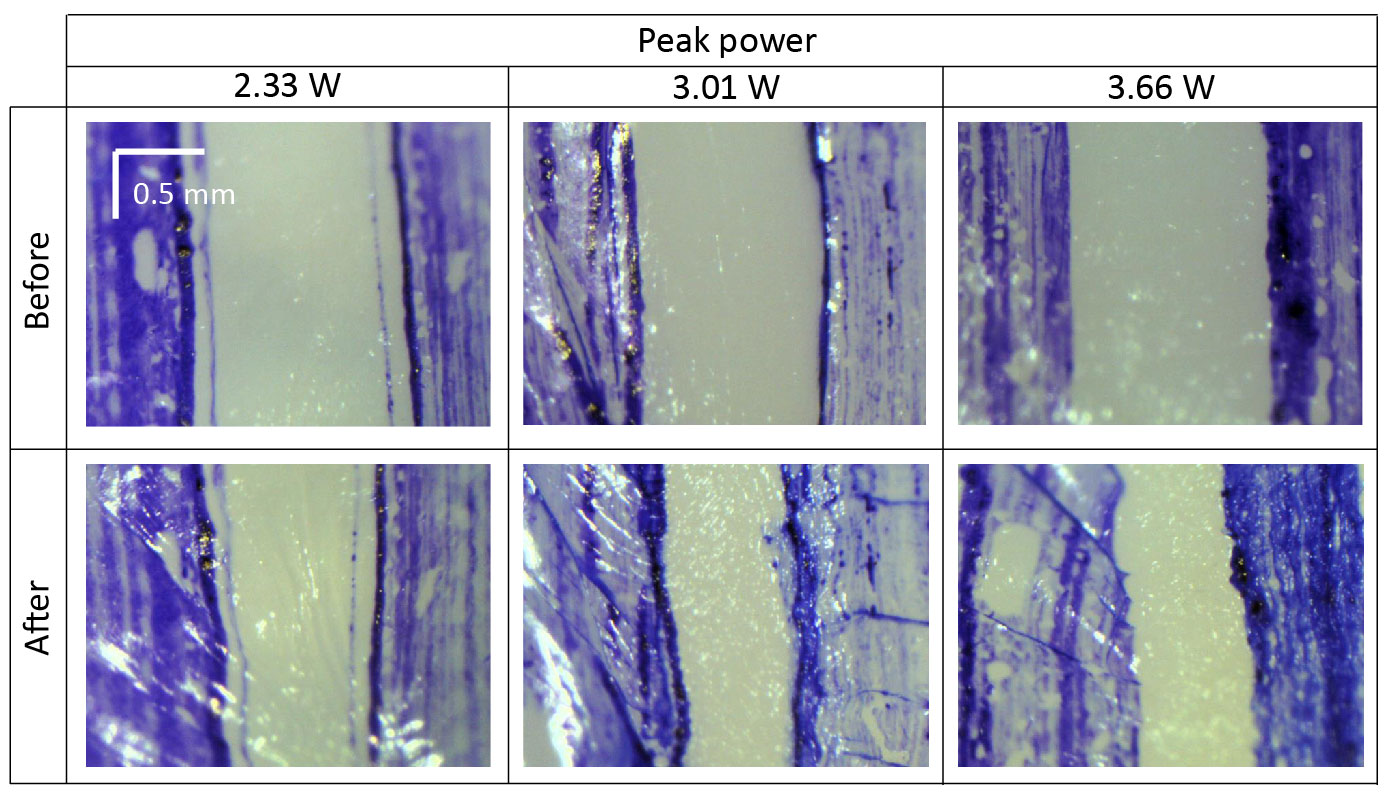}
\caption{Photographs of the regions in the porcine eyes before and after LTC with different laser powers.  The duty cycle is set at 20\%. The blue regions are the marks from the tissue marker. The top row show the photographs before LTC and the corresponding photographs after the LTC are shown in the bottom row.  Shrinkage of the tissue (the regions between the blue marks) could be seen from all the settings.}
\end{figure}
We investigate the influence of different laser parameters on the shrinkage. Figure 3 shows the microscope images of the porcine eyes before and after LTC with the laser powers of 2.33 W, 3.1 W, and 3.66 W. The pulse duty cycle was set to 20\% here. As demonstrated, the shrinkage increases as the laser peak power increases. For the peak power of 2.33 W, the shrinkage is ~ 21\%. As the peak power increases to 3 W and 3.66 W, the shrinkage increased to 36\% and 45\%, respectively.\\
 \begin{figure}[!htb]
\centering
\includegraphics[width=13cm]{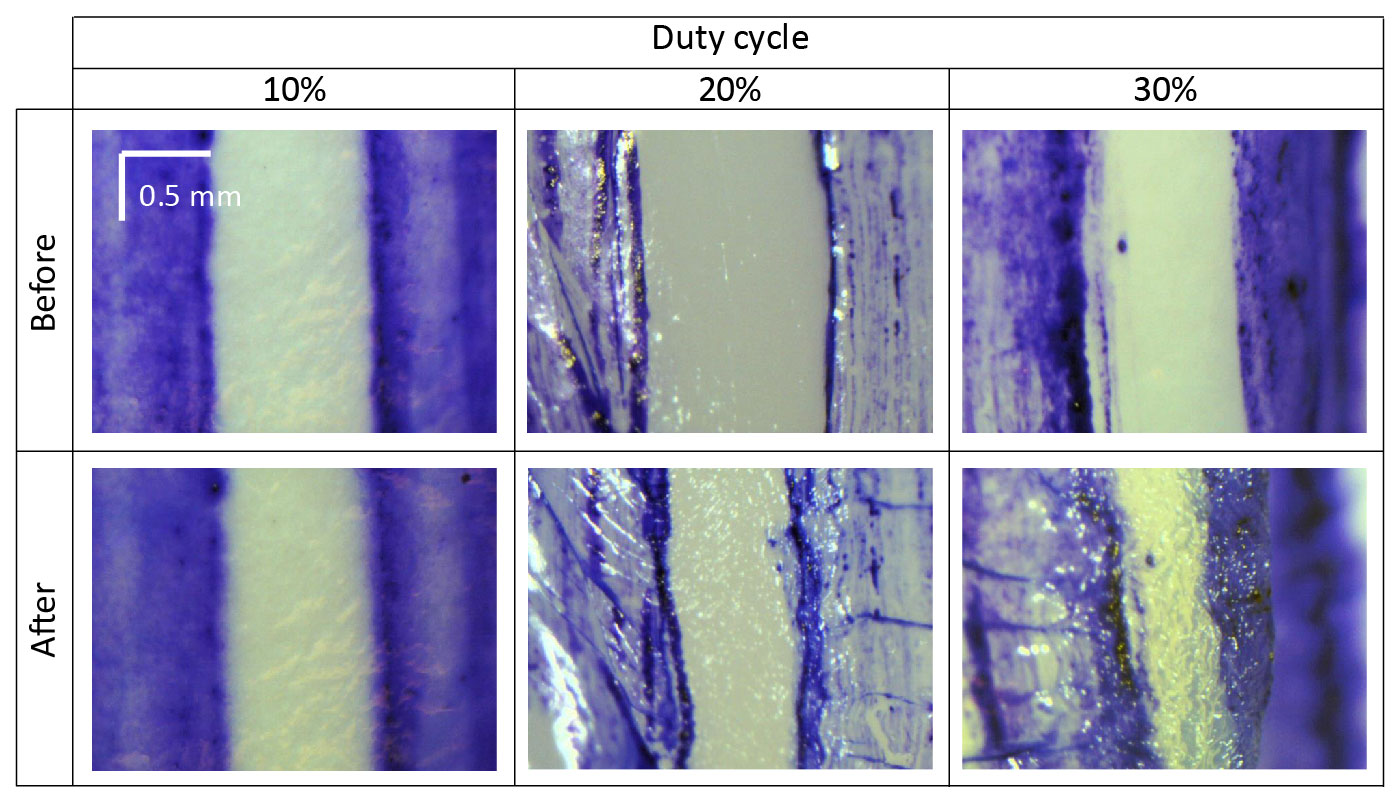}
\caption{Photographs of the porcine eyes before and after LTC with different pulse duty cycle and a peak power of 3W.  The top row show the photographs before LTC and the corresponding photographs after the LTC are shown in the bottom row.}
\end{figure}
\indent From the experiment above, 3 W peak power and 20\% duty cycle is optimum for both the shrinkage and the temperature (the measured temperature is ~ 88 $^{\circ}$C). So we used the 3-W peak power to investigate the influence of the pulse duty cycle on the shrinkage. Figure 4 shows the microscope images of the shrinkage at the pulse duty cycles of 10\%, 20\% and 30\%. It can be seen from the images, at 10\% of the duty cycle, the shrinkage is minimal ($\sim$8\%). When the duty cycle increases to 30\%, the shrinkage is ~ 45\%. However, the tissue surface was damaged due to a high temperature of more than 100 $^{\circ}$C [Fig. 1(b)]. \\
 \begin{figure}[!htb]
\centering
\includegraphics[width=13cm]{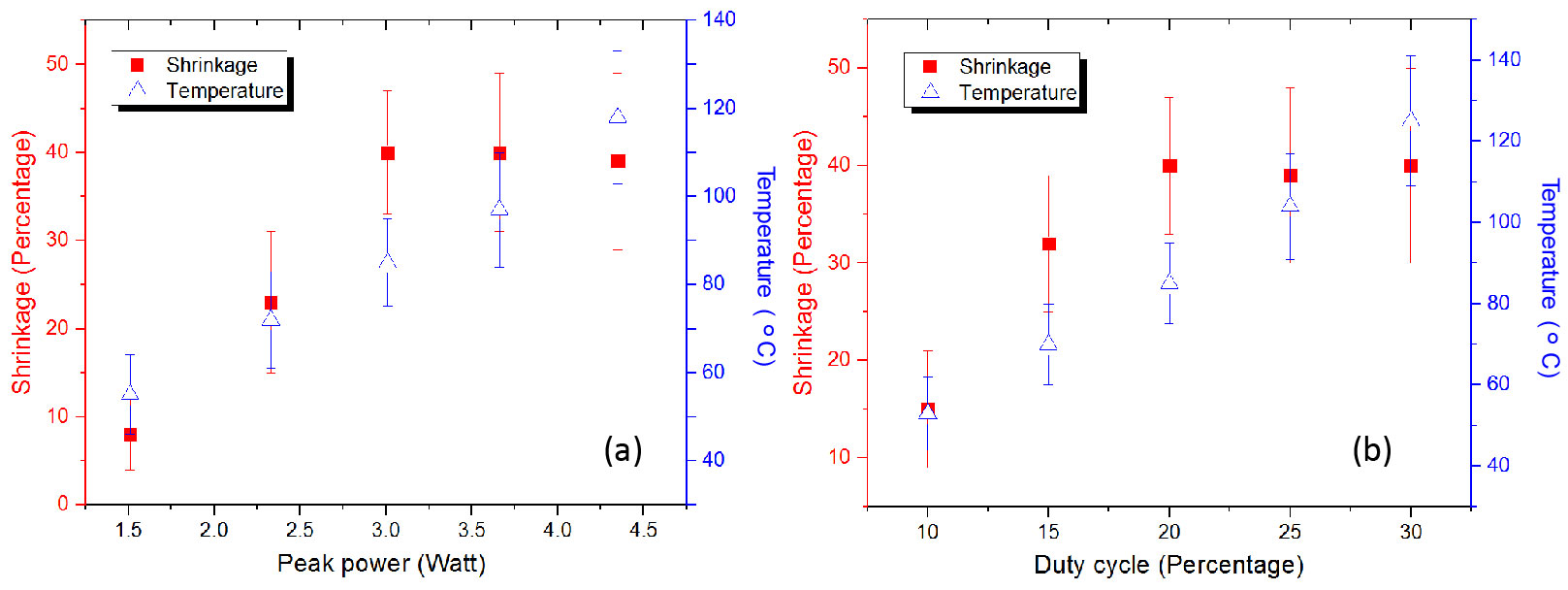}
\caption{(a) Shrinkage and temperature as a function of laser peak power. The pulse laser duty cycle is set at 20\%.   The shrinkage rate saturates with the increase of the peak power and the temperature continues to increase with the increase of the peak power. (b) Shrinkage and temperature as a function of pulse duty cycle. The laser peak power was set as 3 W. The shrinkage rate saturates with the increase of the duty cycle and the temperature continues to increase with the increase of the duty cycle.}
\end{figure}
\indent To determine the optimal laser parameters for the LTC, it is necessary to analyze the relationship between the shrinkage and the tissue temperature at different laser parameters. For a specific laser parameter combination, the shrinkage and temperature was measured 6 -- 8 times on different samples and the averaged values were obtained. Figure 5(a) shows the shrinkage and temperature as a function of laser peak power. The default duty cycle of 20\% was used. It can be seen the temperature keeps increasing as the peak power increases while the shrinkage starts to saturate when the peak power is higher than 3 W. A peak power of 3 W is ideal for the LTC application. The shrinkage and temperature as a function of pulse duty cycle is shown in Fig. 5(b). The peak power was set at 3 W. When the duty cycle is 30\%, the temperature is above 110 $^{\circ}$C while the shrinkage is comparable with that of the 20\% duty cycle.
\subsection*{Shrinkage using angled forceps}
\begin{figure}[!htb]
\centering
\includegraphics[width=13cm]{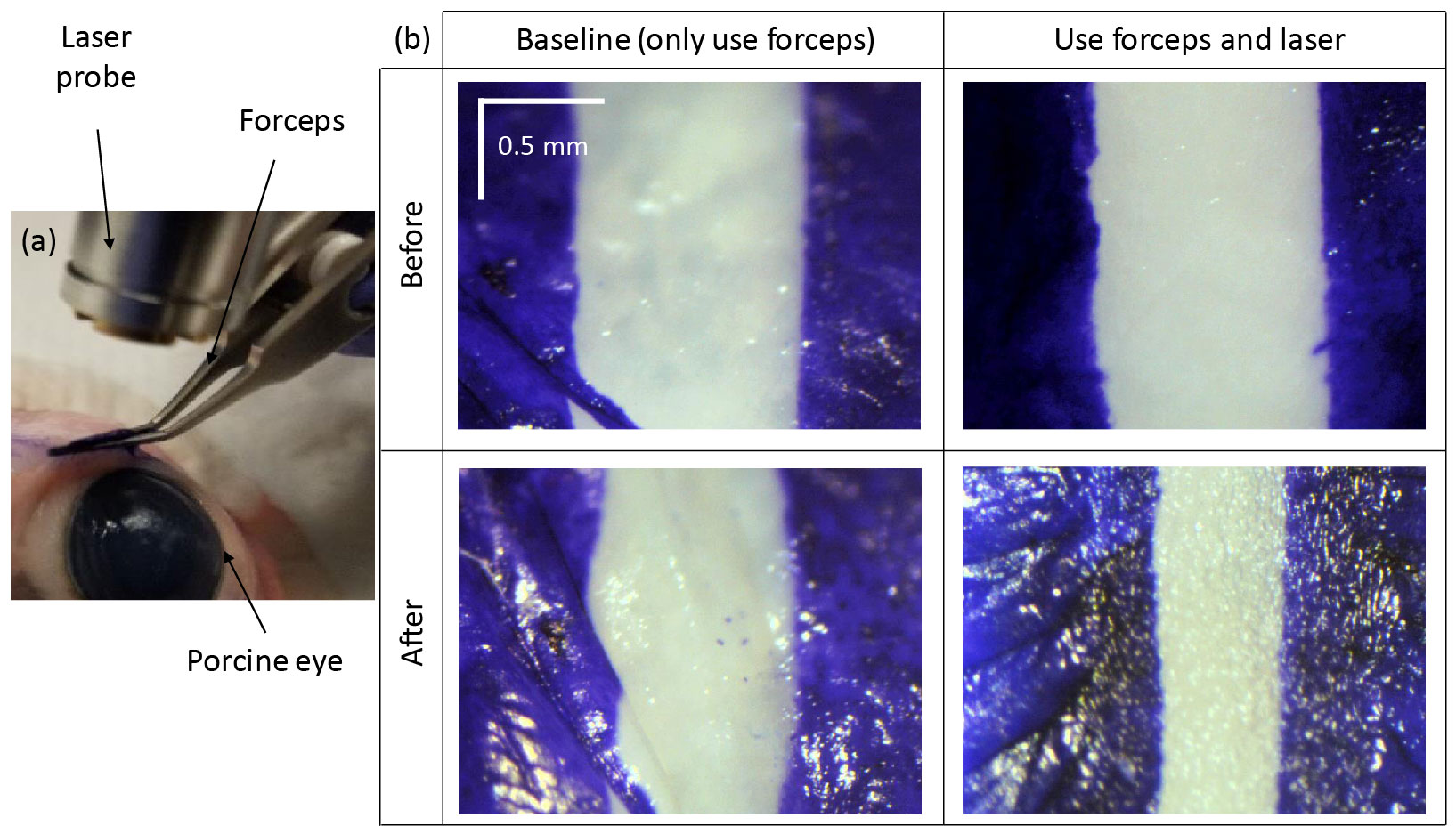}
\caption{(a) Photograph of the LTC setup using the angled forceps. (b) The shrinkage results using only the forceps and both the forceps and the laser.}
\end{figure}
The probe is designed to work with the angled forceps so that the working distance is guaranteed, the shining angle is normal and only the tissue fold held by the forceps is heated. Due to the force from the forceps, the tissue may not be able to recover to its original shape and this may introduce a false shrinkage.  To eliminate this, we test the distance changes between the parallel maker lines due to only the force from the angled forceps. As shown in the left panels of Fig. 6(b), a slight change (~10\%) between the marker lines was found. After the LTC, the shrinkage was measured to be ~ 45\% [right panels of Fig. 6(b)].
\section*{Discussion and conclusion}

The ex vivo porcine eye experiments of the NIR LTC is a proof-of-concept of this new technique. The texture and contents of the conjunctiva of the porcine eye is very close to those of the human eye, so we can expect to achieve similar shrinkage from conjunctivoplasty surgery of human subjects by using this NIR LTC system. However, before testing it on human subjects, further in vivo animal experiments are still needed based on the following considerations. (1) The initial temperature of the porcine eye was about 20 $^{\circ}$C, which is lower than that of the human patients. Therefore, the optimal laser parameters decided here may not be appropriate for the actual treatment of redundant conjunctiva. The laser parameters need to be calibrated by using in vivo animals with a similar body temperature as human being. (2) The ex vivo eye does not have blood circulations, therefore a living eye is needed to assesses conjunctival hyperemia (a measure of inflammation), hemorrhage (bleeding inside tissue), and blanching (a measure of ischemia). (3) The ex vivo eye does not have a healing response, therefore in vivo animal experiment are needed to assess epithelial healing, fibroblast repopulation of the stroma, any scarring formation, and whether the shrinkage is maintained in the long term.\\
\indent In summary, we have developed a NIR LTC using a 1460-nm wavelength laser, which can be efficiently absorbed by water. We developed a compact handheld probe with optimized features for LTC, including a line-shape focus spot and angled forceps. The laser parameters were optimized and ex vivo experiments of on porcine eye was performed. Up to 50\% percent of conjunctiva shrinkage was observed.

\section*{Materials and methods}
\begin{figure}[!htb]
\centering
\includegraphics[width=11cm]{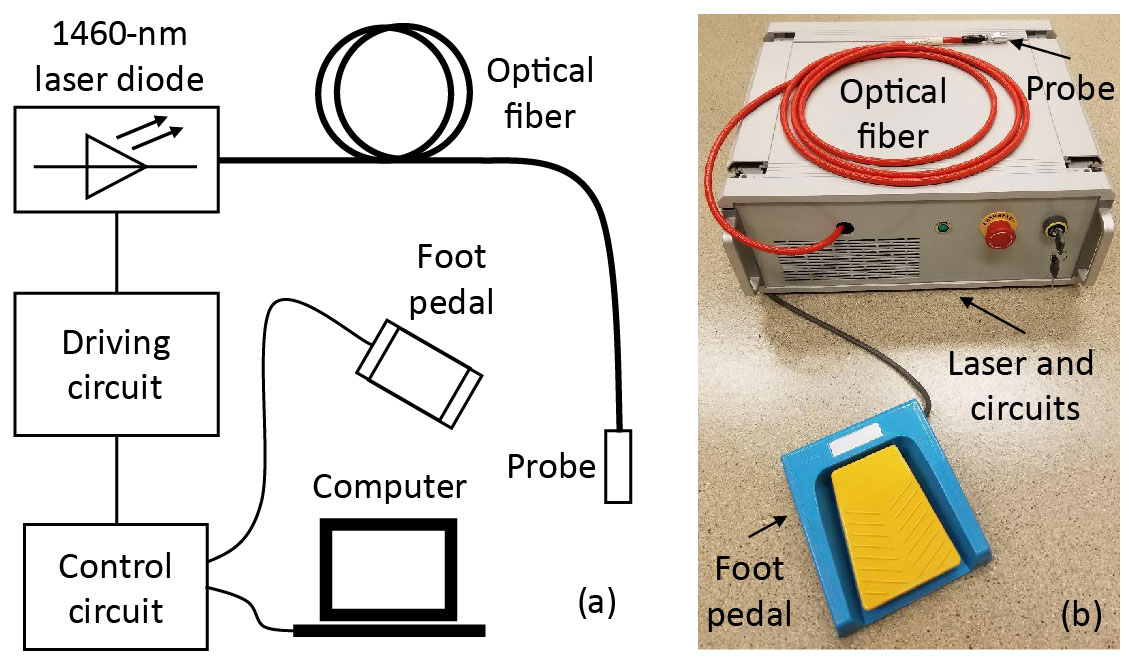}
\caption{(a) Schematic of the 1460-nm LTC laser system. (b) Photograph of the prototype of the LTC laser system.}
\end{figure}
 \begin{figure}[!htb]
\centering
\includegraphics[width=11cm]{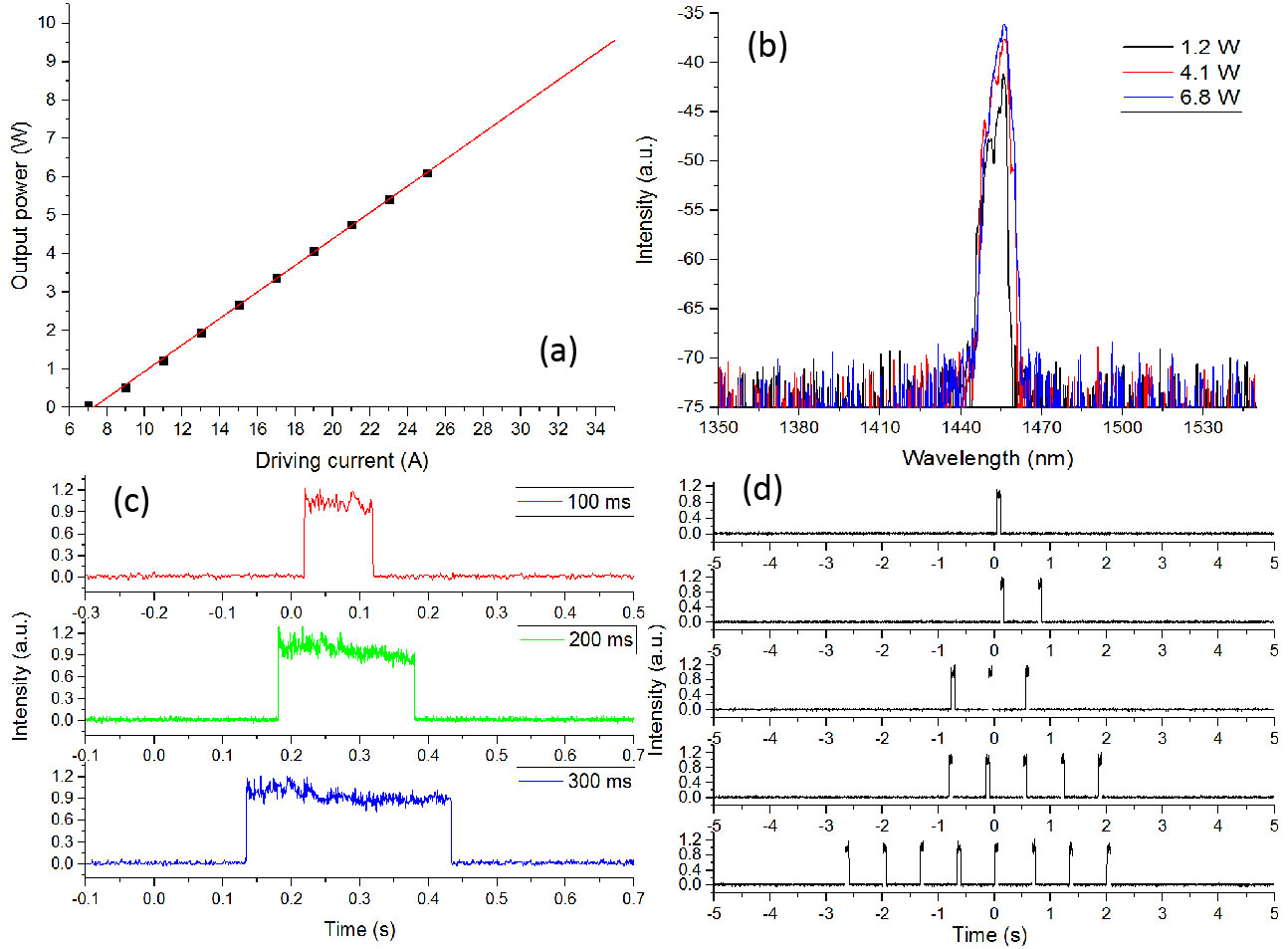}
\caption{(a) Output power of the 1460-nm diode laser is proportional to the driving current. (b) The laser output spectrum at different output powers. (c) Temporal feature of output pulses with different pulse widths. (d) Output pulse trains with different pulse numbers.}
\end{figure}

\subsection*{1460-nm programmable laser diode system}
Figure 7(a) shows the schematic of the programmable pulsed laser diode system for the LTC. A fiber-coupled high-power laser diode module with a maximum continuous-wave (CW) power of 12 W (M1F2S22-1470.10-12C-SS5.x, DILAS, Tucson, AZ, USA) was employed as the light source. The light was output through a multimode fiber with a core diameter of 200 $\mu$m and a numerical aperture (NA) of 0.22. The other end of the fiber was connected to a handheld probe. A 650-nm laser diode was integrated into the source laser for aiming purpose. The power for the aiming light was 200 $\mu$W.\\
\indent Custom built control circuits were used to drive the 1460-nm laser module. The pulse duty cycle, repetition rate, output power and working duration are tunable through a LabVIEW  (LabVIEW 2015, National Instruments, Austin, TX, USA) based control software. The duty cycle, repetition rate, and working duration is arbitrarily tunable. Limited by the capacities of the commercial laser diode module, the maximum output power and the shortest pulse width are 12 W and 100 $\mu$s, respectively. A 650-nm aiming light could be enabled and disabled by the operator from the control software. A foot pedal is used as a trigger for the laser output. Once the laser output is triggered, the 650-nm aiming light will be turned off automatically and the 1460-nm laser will be delivered to the probe according to the preset parameters. However, if the foot pedal is released during the procedure, the infrared laser will be turned off immediately. The photograph of the prototype of the LTC laser system is shown in Fig. 1(b). The whole laser system was assembled into a case with a dimension of 41$\times$36$\times$15 cm3.  \\
\indent The power, spectral, and temporal characteristics of the 1460-nm laser system were measured. The relationship between output power and driving current is shown in Fig. 8(a) and a linear relationship between them could be found. The  lasing threshold current is 6.8 A. The power was measured by a thermal power sensor (S310C, Thorlabs, Newton, NJ, USA).  Figure 8(b) shows the laser spectrum with the output power of 1.2 W, 4.1 W and 6.8 W, respectively. The results were measured by an optical spectrum analyzer (AQ6370C, YOKOGAWA, Tokyo, Japan). The wavelength at the peak power is $\sim$1456 nm. No obvious spectral shift is observed as the output power increases. The spectral bandwidth increases from 3.1 nm at 1.2 W to 3.9 nm at 6.8 W. Compared with the bandwidth of the water absorption peak ($\sim$ 100 nm), the influence of this spectral broadening on the LTC can be neglected. The temporal features of this laser system were measured by a photodetector (PDA10CF, Thorlabs, Newton, NJ, USA) and an oscilloscope (MSO4104B-L, Tektronix, Beaverton, OR, USA). Figure 8(c) demonstrates the temporal shapes of output pulses with the pulse widths of 100 ms, 200 ms, and 300 ms. Figure 8(d) shows the output pulse trains with the pulse numbers following the Fibonacci sequence, which proves the output pulse number can be precisely controlled by the working duration.
\subsection*{Handheld LTC probe}
 \begin{figure}[!htb]
\centering
\includegraphics[width=11cm]{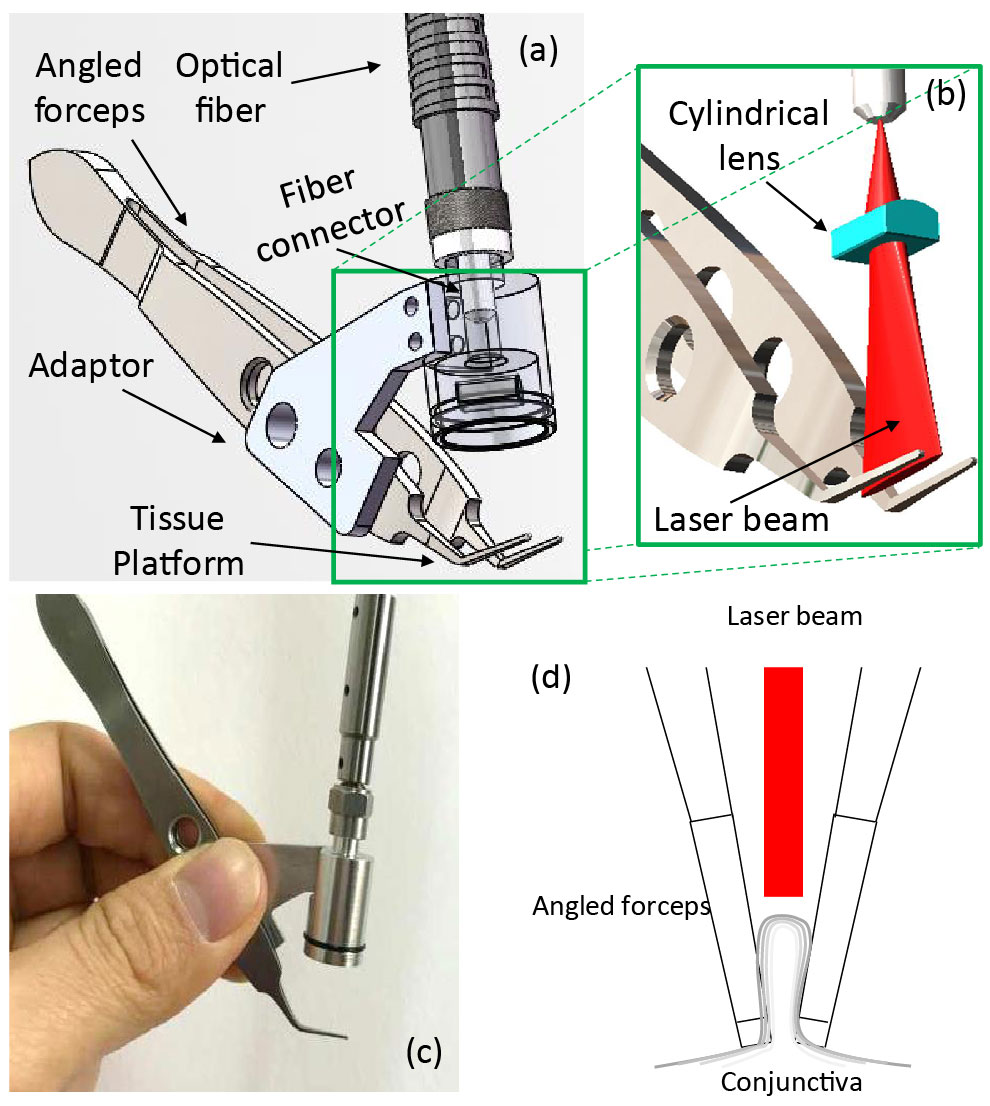}
\caption{(a) Simulation result of the laser beam transformation through the designed optics of the handheld LTC probe. (b) Photograph of the visible pilot laser from the handheld probe. (c) Photograph of the 1460-nm laser on a laser viewing card.}
\end{figure}
 \begin{figure}[!htb]
\centering
\includegraphics[width=11cm]{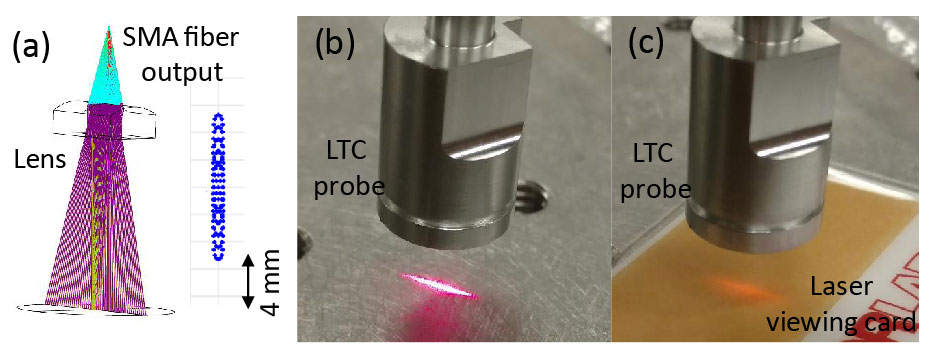}
\caption{(a) 3D model of the handheld probe with the angled forceps. (b) 3D demonstration of the line-focused laser beam and the angled forceps. (c)  Photograph of the handheld LTC probe. (d) Illustration of the configuration for laser delivery to redundant conjunctival tissue.}
\end{figure}

A cylindrical lens is used to focus the SMA fiber output into a line [Fig. 9(b)].  This design reduces the power loss and keeps the probe compact. The ray-tracing (OpticStudio, Zemax, LLC, Seattle, WA, USA) of the light path is shown in the left of Fig. 9(a). The cylindrical lens (LJ1918L1-C, Thorlabs, Newton, NJ., USA) has a clear aperture of 6$\times$4 mm$^2$ and a focal lens of 5.8 mm. The focal plane has dimensions of ~ 10$\times$1 mm2 as shown on the right of Fig. 9(a). Figure 9(b) and 9(c) are the photographs for the probe with the visible aiming light and the 1460-nm laser on, respectively.\\
\indent Figure 10 shows the 3D model and the photography of the handheld probe with the angled forceps.  The probe head and the forceps are connected through an adaptor [Fig. 10(a)].  The line-shaped focal plane is just above the angled platform of the forceps so that the conjunctiva folds grasped by the angled platform can be effectively heated [Fig. 11(b)]. The adaptor also ensures the right angle illumination of the NIR laser light on the conjunctiva fold [Fig. 10(a) and 4(b)] and a proper working distance.  Figure 10(c) shows the actual probe and angled forceps held in hand. Figure 10(d) shows the configuration for laser delivery to redundant conjunctival tissue. A line-focused laser beam is delivered to the conjunctive fold held by a pair of angled forceps. The laser energy is applied in pulses that confine the peak heating to the conjunctival fold. Multiple pulses are delivered to achieve collagen shrinkage, which can be directly visualized by the surgeon.
\subsection*{Laser parameter selection}
A pulsed laser will be used to heat and shrink the conjunctiva.  The laser beam will be focused into a line on the tissue and the situation can be modeled as a one-dimensional heat diffusion problem for the calculation of tissue thermal relaxation time \cite{niemz96}:
\begin{equation}
\tau = \frac{d^2}{4D}
\end{equation}
where $\tau$ is the tissue thermal relaxation time, $D$ is the heat diffusivity of tissue and $d$ is the heat diffusive length. The heat diffusivity of tissue is approximately 1.3$\times$10$^{-7}$ m$^2$ s$^{-1}$ \cite{tele01}. The thickness of the human conjunctiva is approximately 0.24 mm, and loose conjunctiva folded over when grasped by the surgical forceps should be approximately 0.5 mm thick. For this thickness, the thermal relaxation time is ~ 0.5 s. So the pulse duration needs to be shorter than 0.5 s for the thermal energy to be confined to within this depth.\\
\indent In addition, instead of using a single pulse, we propose to use the cumulative effect of multiple pulses. This strategy can not only reduce the requirement of peak power for effective shrinkage of the conjunctiva, but also keep the optimal temperature for a suitable period of time. The pulse train is emitted at a fixed repetition rate. This repetition rate can be estimated as following. For a pulse period including the pulse-tissue interaction and laser-off thermal relaxation, a thermal relaxation of ~ 0.5 s indicating the pulse duration should be shorter than 0.5 s for confine the thermal energy in depth, while the laser-off duration should be around or a little bit longer than 0.5 s for effective heat dissipation of this pulse. So the pulse period should be ~ 1s, corresponding to a repetition rate of ~ 1 Hz. The effectiveness of this estimation has been validated by the temperature variation in Section 3.1. We set a time duration of 4 seconds for the LTC process, which corresponds to 4 pulses at a repetition rate of 1 Hz.
\subsection*{Experiment setup and procedure}
 \begin{figure}[!htb]
\centering
\includegraphics[width=11cm]{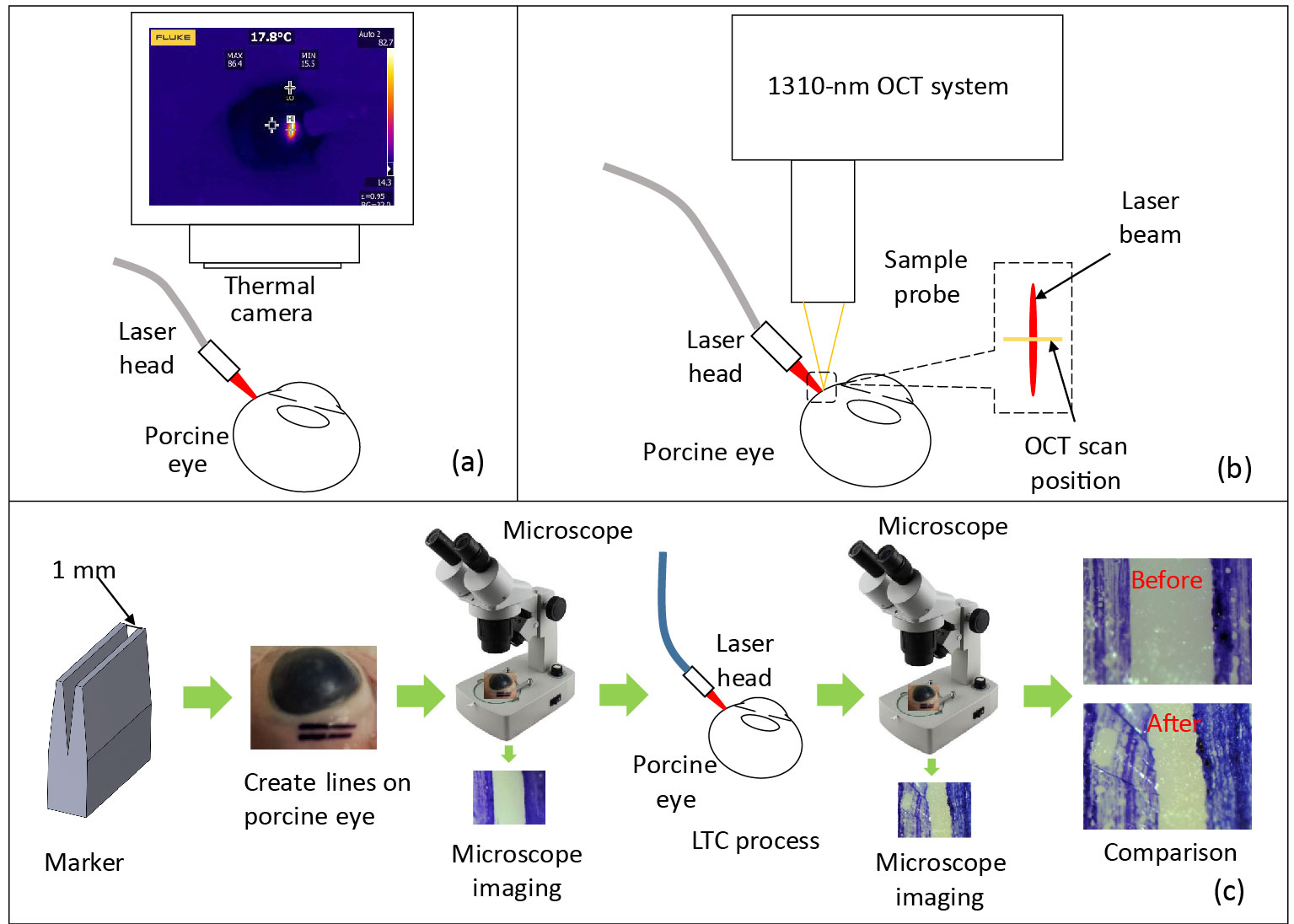}
\caption{(a) Setup of measuring laser-induced temperature variation by a thermal camera. (b) Schematic of recording the dynamic shrinkage process by a 1310-nm optical coherence tomography system. (c) Procedure to measure the shrinkage from the LTC.}
\end{figure}
Figure 11 shows the setup and procedure of the experiment. Porcine eyes (Animal Technologies, Tyler, TX, USA) were used as samples in this LTC study. Frozen porcine eye balls were thawed in phosphate buffered saline before the experiment. To keep the condition of the tissue consistent, we used each position on the conjunctiva once. Because each porcine eye has a finite area of the conjunctiva and we were using a beam size of 10$\times$1 mm$^2$, the laser radiation on a single eye was limited to 6 – 8 times. Around 30 porcine eyes were used in this study. A 3D-printed marker was used to create a specific, observable region of interest that was denoted by two parallel lines on the conjunctiva surface: blue colors in the microscope images in Fig. 11(c) with a length of 10 mm and a separation of 1 mm.  The line-focused light from the LTC probe was targeted to shine between the parallel marker lines on the conjunctiva.  The angled forceps which grasp the tissue may introduce false shrinkage due to force applied on the conjunctiva. So, the angle forceps will not be used in the experiments for optimizing the laser parameters. A thermal camera (TiS45, Fluke, Everett, WA, USA) was used to record the temporal temperature change in the region during the LTC. OCT was also used to monitor the tissue structural change during the LTC.  Figure 11(a) shows the experimental setup for LTC (bottom) and thermal camera image (top) of the porcine eye during the LTC. The thermal camera allows the record of the images during the LTC. A custom 1310-nm swept-source OCT system with an A-line rate of 50 kHz, a transverse resolution of 15 $\mu$m, and an axial resolution of 8.5 $\mu$m was used to monitor the LTC process. The detailed information about the OCT system has been described in a previous publication \cite{su15}. Figure 11(b) shows the schematic of the OCT experiment. The OCT B-scan covered a range of 5 mm and repeated B-scan at this same location was recorded. The OCT imaging rate is 50 frames per second. The laser-induced shrinkage was measured [Fig. 11(c)]. The photographs of the targeted area before and after the LTC were taken with a digital microscope (OMAX, Gyeonggi-do, Korea).  The shrinkage percentage of the region was obtained by comparing the two images.

\section*{Acknowledgements}
We would like to thank Wen Chen from TXStar Laser Technology (Shanghai) Co., Ltd and Minzhao Zhang from Shanghai CONSUN Photoelectric Technology Co., Ltd for helping build the laser system and Chao Pang from Heraeus Noblelight for fabricating the handheld probe.

\section*{Funding}
National Institutes of Health (R01 EY023285, R01 EY018184); Oregon Health and Science University Foundation; Oregon Clinical \& Translational Research Institute (Biomedical Innovation Program).
\section*{Financial disclosures}
Oregon Health \& Science University (OHSU), Baylor College of Medicine, David Huang, Gangjun Liu, Jianlong Yang, and Stephen C. Pflugfelder have potential financial interest in pending patent of the device described in the paper. These potential conflicts of interest have been reviewed and managed by OHSU.
\section*{Competing Financial Interests}
The authors declare that they have no competing interests.
\section*{Author contributions statement}
G.L., D.H., J.Y., and S.C.P. designed the experiment. J.Y. and G.L. wrote the manuscript. J.Y., R.C., V.G., and T.B. conducted the experiments. G.L., D.H., and S.C.P. supervised the project. All the authors discussed the results and commented on the manuscript.

%\section*{Additional information}

%To include, in this order: \textbf{Accession codes} (where applicable); \textbf{Competing financial interests} (mandatory statement). 

% The corresponding author is responsible for submitting a \href{http://www.nature.com/srep/policies/index.html#competing}{competing financial interests statement} on behalf of all authors of the paper. This statement must be included in the submitted article file.

\end{document}